\documentclass[preprints,article,accept,moreauthors,pdftex]{Definitions/mdpi} 
\firstpage{1} 
\pubvolume{1}
\issuenum{1}
\articlenumber{0}
\pubyear{2023}
\copyrightyear{2023}
\externaleditor{Academic Editor: Lev Vaidman}
\datereceived{14 February 2023} 
\daterevised{17 April 2023}
\dateaccepted{22 April 2023} 
\datepublished{} 
\hreflink{https://doi.org/} 

\usepackage{amssymb}

\usepackage[english]{babel}

\newcommand{\ignore}[1]{}

\usepackage{color}

\newcommand{\beq}{\begin{equation}}
\newcommand{\eeq}{\end{equation}}
\newcommand{\bdm}{\begin{displaymath}}
\newcommand{\edm}{\end{displaymath}}
\newcommand{\bea}{\begin{eqnarray}}
\newcommand{\eea}{\end{eqnarray}}

\newcommand{\benum}{\begin{enumerate}}
\newcommand{\eenum}{\end{enumerate}}
\newcommand{\bit}{\begin{itemize}}
\newcommand{\eit}{\end{itemize}}
\newcommand{\bdes}{\begin{description}}
\newcommand{\edes}{\end{description}}
\newcommand{\bpic}{\begin{picture}}
\newcommand{\epic}{\end{picture}}
\newcommand{\bc}{\begin{center}}
\newcommand{\ec}{\end{center}}

\newcommand{\bq}{\begin{quote}}
\newcommand{\eq}{\end{quote}}

\Title{The Open Systems View and the Everett Interpretation}

\TitleCitation{The Open Systems View and the Everett Interpretation}

\Author{Michael 
 E. 
 Cuffaro *
 $^{,\dagger,\ddagger,}$\orcidA{} and Stephan Hartmann $^{\ddagger}$\orcidB{}}

\AuthorNames{Michael E. Cuffaro, Stephan Hartmann}

\AuthorCitation{Cuffaro, M.E.; Hartmann, S.}

\address[1]{%
Munich Center for Mathematical Philosophy, Ludwig Maximilian University Munich, 80539 Munich, Germany}

\corres{Correspondence: michael.cuffaro@lmu.de}

\firstnote{Current address
: Geschwister-Scholl-Platz 1, 80539 Munich, Germany}
\secondnote{These authors contributed equally to this work.}

\abstract{It is argued that those who defend the Everett, or `Many Worlds', interpretation of quantum mechanics should embrace what we call the \emph{general quantum theory of open systems} (GT) as the proper framework in which to conduct foundational and philosophical investigation in quantum physics. GT is a wider dynamical framework than its alternative, \emph{standard quantum theory} (ST). This is true even though GT makes no modifications to the quantum formalism. GT rather takes a different view, what we call \emph{the open systems view}, of the formalism; i.e., in GT the dynamics of systems, whose physical states are fundamentally represented by density operators, are represented as fundamentally open as specified by an in general non-unitary dynamical map. This includes, in principle, the dynamics of the universe as a whole. We argue that the more general dynamics describable in GT can be physically motivated, that there is as much \emph{prima facie} empirical support for GT as there is for ST, and that GT could be fully in the spirit of the Everett interpretation---that there might, in short, be little reason for an Everettian not to embrace the more general theoretical landscape that GT allows one to explore.}


\begin{document}

\section{Introduction}
\label{s:intro}

The so-called `measurement problem' of quantum theory, as it is commonly framed (see, e.g., \citep[]{myrvoldInFreireVolume}, p. 63),\footnote{This is not the only way to approach the question of how to make sense of measurement in quantum theory. Another approach, which allows one to make finer distinctions among the various interpretations of the formalism, identifies two separate problems: (a) the problem to account for the, according to quantum theory, objective indeterminacy associated with a given measurement context, and (b) the problem to account for the mutual incompatibilities that exist between the various possible measurement contexts associated with a given dynamical system. For more on this, see \citep[]{3m2020}, pp. 10--12, 223--225. For our purposes it will be sufficient to frame the measurement problem in the more traditional form.} runs as follows. On the one hand, in the absence of a measurement, the dynamics of a given quantum system are unitary according to the theory. On the other hand, given a measurement, the apparent `collapse' of the vector representing a system's dynamical state, in accordance with the Born rule, is in general non-unitary. Positing, via the Born rule, a special measurement dynamics over and above the standard unitary dynamics is widely regarded as ad hoc, however. After all, from a physical point of view, measurement interactions are just dynamical interactions like any other. In a sense this is true even according to quantum theory, since a quantum description of a measurement interaction can be given in unitary terms to any desired level of detail given an appropriate placement of the so-called `Heisenberg cut', on one side of which lies our quantum description of a measurement interaction, and on the other our classical description of our observation of its result. Stated in these terms, the measurement problem is that of closing this gap, either by proposing a new theory to explain the connection between them, or by showing that, upon reflection, it is already fully explained by quantum theory. The Everett, or `Many Worlds', interpretation of quantum theory chooses the second option.\footnote{Everett is not the only interpretation that chooses not to supplement the theory. For examples of other contemporary interpretations of this kind, see \citep[]{brukner2017, bub2016, bub-pitowsky2010, demopoulosOnTheories, fuchsEtAl2014, healey2017, 3m2020, rovelli2021}. For further discussion of such interpretations in the context of open systems, see \citep[]{cuffaroHartmannOpenSystemsView}.}

In the form of quantum theory presented in most textbooks on the topic---what we will be calling \emph{standard quantum theory} (ST)---the physical state of a system, $\mathcal{S}$, at a time $t$ is represented by a normalized state vector, $| \psi(t) \rangle$; one element in a Hilbert space, $\mathcal{H}_{\mathcal{S}}$, representing $\mathcal{S}$'s possible states. $\mathcal{S}$'s dynamical evolution is given by
\begin{align}
  \label{eqn:unitary_st}
  | \psi (t) \rangle = U(t) \, | \psi (0)\rangle,
\end{align}
where the time evolution operator $U(t)$ $=$ $\exp (- {\rm i} \, H \, t)$, with $H$ a Hermitian operator representing the system's Hamiltonian whose eigenvalues are the possible energy values of the system.

ST is not itself a particular physical theory. It is an abstract theoretical framework within which various particular relativistic and non-relativistic quantum theories may be expressed \citep[]{cuffaroHartmannOpenSystemsView, 3m2020, nielsenChuang2000, wallace2019}. The particular form of the dynamics of a given class of systems is the purview of the particular quantum theory that pertains to that class. But there are a number of things that can be said about every dynamical model of ST irrespective of the particular quantum theory that model has been formulated in. Here we note the following two. (I) It follows from the above assumptions that $U(t)$ is a unitary map on the state space of $\mathcal{S}$. (II) $S$ is a closed system. This follows from the fact that $H$, which describes the possible energy values of the system, includes no terms representing its interaction with an environment. We will call any theoretical framework, like ST, in which all phenomena are modeled as closed systems a framework formulated in accordance with the \emph{closed systems view} \citep[]{cuffaroHartmannOpenSystemsView}.

ST is a highly successful theoretical framework that has been used to study all sorts of phenomena, even though many of these---so-called `open systems phenomena'---cannot be modeled in a simple way as closed systems. Lasers, which are pumped by an external energy source, are an example. The spontaneous emission of a photon by an atom is another. In cases like these one models the phenomena in terms of a dynamical coupling between the system of interest and an idealized representation of its environment, $\mathcal{E}$, such that the combined state of $\mathcal{S+E}$, which will be entangled in general (although one usually assumes that $\mathcal{S}$ and $\mathcal{E}$ are initially uncoupled or only weakly coupled), is given by the state vector $| \Psi^{\mathcal{S+E}} \rangle$.

It is not possible to represent the state of a subsystem, like $\mathcal{S}$, of an entangled system using a state vector in ST. However there is a probabilistic generalization of the state vector,
\begin{align}
  \label{eqn:density_operator}
  \rho = \sum_i p_i| \psi_i \rangle\langle \psi_i |,
\end{align}
where $\rho$ is called the \emph{density operator} associated with the system and the $p_i$ are interpreted as probabilities with $\sum_i p_i = 1$, that can be used to represent $\mathcal{S}$'s probabilistic state; though it will not, in this case, be possible to interpret Eq. \eqref{eqn:density_operator} literally, i.e., as representing that $\mathcal{S}$ is in the `pure state' $| \psi_i \rangle\langle \psi_i |$ corresponding to the state vector $| \psi_i \rangle$ with probability $p_i$ (since a subsystem of an entangled system can never be in a pure state). For the purposes of providing an in general probabilistic description of the dynamical phenomena that we associate with $\mathcal{S}$, however, the density operator completely suffices in the sense that every probability measure over yes-or-no questions concerning the observable quantities associated with any system is representable by means of a density operator acting on that system's state space \citep{gleason1957}.\footnote{Gleason's theorem \citep[]{gleason1957} represents measurements as projections and is valid for Hilbert spaces of dimension $\geq 3$. See \citet[]{busch2003} for a generalization.}

Given that $\mathcal{S}$'s initial (probabilistic) state is described by the density operator $\rho^{\mathcal{S}}(0)$, to derive the (generalized) dynamics of $\mathcal{S}$ up to some time $t$ one first calculates the dynamical evolution of the combined state, $| \Psi^{\mathcal{S+E}} \rangle$, from the initial time until $t$ (in the case of non-relativistic quantum theory using the Schr\"odinger equation, for instance). One then takes the partial trace over the combined state with respect to $\mathcal{E}$, yielding a reduced density operator, $\rho^{\mathcal{S}}(t)$, representing $\mathcal{S}$'s final (probabilistic) state. The above procedure may also be effectively and conveniently represented as an in general non-unitary dynamical map, $\Lambda$, mapping density operators in the probabilistic state space of $\mathcal{S}$ to other density operators, though it must not be forgotten that such non-unitary evolution is always only effective in ST; since a system's dynamics is always modeled as generated by its associated Hamiltonian and is unitary, it follows that non-unitary dynamics must always be taken to be a mere manifestation of an underlying unitary dynamics, in this case of the larger system $\mathcal{S+E}$.\footnote{\citet{cuffaroHartmannOpenSystemsView}, from whom we have taken the terminology ``ST'' and ``GT'' (to be introduced in the next section), discuss a number of different senses of fundamentality relevant to the concepts of a given theoretical framework, as well as between frameworks. What we mean by fundamental here is what they call OntFund-O, the relation of `ontic fundamentality' within a given framework. A given concept is fundamental, in that sense, in a given framework if and only if instances of the concept are not described in the framework as determined by anything else.}

\section{The general quantum theory of open systems}
\label{s:gt}

ST is not the only theoretical framework that can be formulated using the quantum-mechanical formalism. There is an alternative, \emph{the general quantum theory of open systems} (GT), that takes a different view---the \emph{open systems view}---of the formalism. On the open systems view, the dynamics of interacting systems, not the dynamics of closed systems, are taken to be fundamental, so that the influence of a system's environment is not represented in terms of a dynamical coupling between two parts of one closed system, but fundamentally via the equations that govern the dynamics of the system of interest. Accordingly, in GT a physical system, $\mathcal{S}$, is fundamentally represented in terms of an in general non-unitarily evolving density operator whose dynamics is governed by a linear (in the sense of preserving mixtures), positive, trace-preserving dynamical map, $\Lambda$, acting on $\mathcal{S}$'s state space.\footnote{Note that GT is, essentially, the framework originally laid out in \citep{sudarshan1961, jordan1961}; and subsequently elaborated upon in, for instance, \citep[][]{jordan2004, shaji2005}; as well as in the works of other authors. For more on the properties of not completely positive maps than we will be discussing below, for instance, see \citep[]{dominyEtAl2016}.}

Although both employ the same quantum formalism, GT is a wider dynamical framework than ST.\footnote{GT is a theoretical framework that allows for the modeling of more general dynamic evolutions of a system than ST. However, which dynamics one chooses depends on the problem at hand and on the physical principles one wishes to assume in the case under study. Dynamical collapse theories, for example, assume that collapse occurs in position space, and the resulting dynamics can be derived from a suitable Lindblad equation. Collapse theories are of course controversial and they run into problems with a relativistic extension. However an advocate for GT is not obliged to accept collapse theories. It is merely possible to formulate them within this framework.} The reason is that, although GT requires that $\Lambda$ be a positive map on the valid states of $\mathcal{S}$, unlike in ST it is not required in GT that $\Lambda$ be completely positive in a sense we will presently explicate. The standard argument for imposing what is called the `principle of complete positivity' is that we should require the effect of a given map, $\Lambda$, to be a valid dynamical evolution for every possible initial state of $\mathcal{S}$ irrespective of the existence of a `witness' system, $\mathcal{W}_n$, of a given dimensionality $n$ that is not interacting with $\mathcal{S}$ (though it may have in the past). The trouble, according to this argument, with a positive but not completely positive map, $\Lambda$, on $\mathcal{S}$'s state space is that extending $\Lambda$ to include the, let us assume, trivial dynamics of the witness: $\Lambda \otimes I_n$, will in general result in a final state that yields negative probabilities for the outcomes of certain measurements on $\mathcal{S}+\mathcal{W}_n$. Indeed, requiring that a map, $\Lambda$, be completely positive on its full state space is equivalent to requiring that $\Lambda \otimes I_n$ be a positive map for all $n$ and the principle is usually defined explicitly in this way (see, e.g., \citep[]{breuer2007}, p. 86). The standard argument is less than compelling, however. Note first that a prediction of negative probabilities for certain measurements is only possible if $\mathcal{S}$ is initially entangled with $\mathcal{W}_n$.\footnote{Our reply to the standard argument is drawn mainly from \citep[]{shaji2005}.} But in that case $\mathcal{W}_n$ should really be considered to be a part of $\mathcal{S}$'s environment, $\mathcal{E}$; and it can be shown that, in fact, there actually is no completely positive map that can describe the dynamics of $\mathcal{S}$ when it is initially entangled with $\mathcal{E}$ (\citep[]{jordan2004}, pp. 13--14).\footnote{\citeauthor{jordan2004}'s result generalizes an earlier result for two-dimensional systems proved by \citet[]{pechukas1994}.} We should not trouble ourselves about this too much, however, for if $\mathcal{S}$ is a subsystem of an entangled system then certain states of $\mathcal{S}$ will be impossible; for instance, pure states, or generally any state that is not a valid partial trace over the entangled state of the overall system. With this in mind one can then define a `not completely positive map' that is completely positive \emph{vis \'a vis} the states that are not ruled out by a given setup, while for states that are ruled out we allow that the map may not even be positive on $\mathcal{S}$, let alone its trivial extension positive on $\mathcal{S+W}_n$ (for discussion, see \citep{cuffaroMyrvold2013}).

In the context of ST, one can give a better argument for imposing complete positivity as a dynamical principle, namely that complete positivity is assumed in the derivation of Stinespring's dilation theorem \citep[]{stinespring1955}, which asserts that corresponding to a given $\rho^{\mathcal{S}}$ there is a unique (up to unitary equivalence) pure state $| \Psi^{\mathcal{S+A}} \rangle$ of a larger system $\mathcal{S+A}$ (where $\mathcal{A}$ is called the `ancilla' subsystem), whose dynamics is unitary, and from which we can derive the in general non-unitary dynamics of $\mathcal{S}$. In other words the procedure for deriving the effective dynamics of an open system that we discussed in the last section, i.e., the procedure of first considering the dynamics of the closed system $\mathcal{S+E}$ and then taking the partial trace over its final state with respect to $\mathcal{E}$, is predicated upon the assumption of complete positivity. Since ST is formulated in accordance with the closed systems view, every system is modeled as a part of some closed system by definition, from which it follows that complete positivity must be imposed as a fundamental physical principle. As \citet{raggio1982} put it:

\begin{quote}
A system-theoretic description of an open system has to be considered as phenomenological; \emph{the requirement that it should be derivable from the fundamental automorphic dynamics of a closed system} implies that the dynamical map of an open system has to be completely positive (p. 435, our emphasis).
\end{quote}

Models formulated in GT are under no such restriction. In a framework in which the dynamics of interacting systems are represented as fundamental, there is no need to be able to derive the dynamics of a system in this way. We can in principle describe the dynamics of any system, even the universe as a whole, \emph{as if} it were initially a subsystem of an entangled system.

\section{The open systems view and the Everett interpretation}
\label{s:everett}

Unitarity is clearly deemed to be important by many Everettians. Simon Saunders, for instance, in the context of a discussion of the metaphysics of personal identity, writes that one of the drawbacks of adopting the rule that `persons' and `things' correspond to `branch parts' is that

\begin{quote}
  [i]nvoking it seems to compromise a chief selling point of the Everett interpretation, which is that many-worlds follows from the unitary dynamics, with no added principles or special assumptions. This is what puts the Everett interpretation in a class of its own when it comes to the quantum realism problem: there are plenty of avenues for obtaining (at least non-relativistic) one-world theories if we are prepared to violate this precept. (\citep[]{saunders2011}, p. 193)
\end{quote}

We will not comment on the efforts of Everettians to make sense of the metaphysics of personal identity, nor on the wider context of Saunders' discussion: chance.\footnote{Although we will not comment on these issues, we do think it would be interesting for Everettians to reconsider them in the light of a framework in which fundamental non-unitary dynamics is not a priori ruled out. For instance if the only arguments against certain characterizations of personal identity (like the rule that `persons' and `things' correspond to `branch parts') or of chance are that they are in tension with quantum theory's restriction against fundamental non-unitary evolution, then we believe that Everettians should rethink those arguments. These are not the only issues that may be clarified in a framework in which the dynamics of density operators are taken as basic. \citet[]{chenTimeAsymmetricBJPS, chenNatureOfPhysicalLaws, chen2020}, for instance, discusses the issue of the arrow of time. See also \citet{maroneyGibbsPhysicalBasis} and \citet{robertsonHolyGrail} for related discussions of how taking the density matrix seriously as a real representation of the state of a system helps to clarify the physical basis of the Gibbs entropy and its role in providing a statistical-mechanical underpinning for the second law of thermodynamics.} We merely note that modifying the quantum formalism is only one way to allow for fundamentally non-unitary dynamics. Another way is to recognize that there is more than one way to view the formalism, and that we are only required to regard non-unitary evolution as non-fundamental in a framework that has been formulated in accordance with the closed systems view. If, instead, we take the dynamics of open systems---ultimately all that we really have empirical access to in any case---to be fundamental, then it is clear that such dynamics may in general be non-unitary according to quantum theory.

The density operator representing the state of a subsystem of an entangled system is an objective description of the degrees of freedom of the subsystem being modeled and in that sense it is no less a description of something real in the world than the state of the entangled system as a whole \citep[]{wallaceTimpson2010}. There is, further, nothing conceptually incoherent about describing even the state of the universe as a whole in these terms on the Everett interpretation; this is a description of the universe that David Wallace, for instance, explicitly entertains, at least as a real possibility (see \citep[]{wallace2012}, sec. 10.5).\footnote{The backdrop to Wallace's discussion in \citep[]{wallace2012} is the `black hole information loss paradox'. Wallace's opinion about whether the paradox is evidence for fundamental non-unitarity has shifted over the years (see, e.g., \citep[]{wallace2020a}).} For our part, we think that although it is clear that, empirically, ST is a highly successful---arguably physics' most successful---theoretical framework, there are nevertheless reasons coming from both non-quantum physics as well as from ST's own applications to motivate asking the question of whether it really is as expressive as it needs to be to make sense of the physical world.

Although standard cosmological models based on the FLRW solutions to the Einstein field equation describe a closed universe, they are well-known to be based on strong idealisations introduced with little other reason than that they simplify the relevant mathematics \citep[sec. 1.1]{smeenkEllisCosmologySEP}. The scale factor, for instance, represents nothing physically significant in itself \citep[]{sloan2021}. Hawking's proposal \citep[]{hawking1976b} to model the quantum description of a black hole in terms of a linear map on the space of density operators is, to be sure, controversial (for discussion, see \citep{giddings2013, pageUniverseOpenSystem, wallace2020a}). But at least part of the motivation for wanting to reject it seems to amount to nothing more than that the proposal runs counter to ST (see, e.g., \citep[pp.\ 32--34]{giddings2013}; cf. \citep[p. 219]{wallace2020a}), according to which non-unitary dynamics simply cannot be fundamental. Hawking's particular proposals aside, we are rather inclined to take the opposite view, and to ask the question of whether a closed-system description of the universe is really apt (see also \citep[][]{grybSloanScaleSurplus, sloan2018}).

Against this it will of course be argued that no contradictions have been demonstrated between any of the phenomena currently known and their theoretical descriptions in ST; and since ST is highly empirically successful we should therefore seek to conform whatever picture of nature is suggested by these phenomena to it rather than the other way around. Even putting to one side the question of what fundamental unitary evolution is supposed to mean in the context of the quantization of gravity \citep[]{castroEtAl2020, giacominiEtAl2019, oritiTimeEmergence}, however, this objection misconstrues what the empirical success of ST actually rests upon, namely its applications to systems that are empirically accessible to us. These are the subsystems of the universe. And since neither gravity \citep[]{zeh2007} nor entanglement \citep[]{herbstEtAl2015forQuantumReports, yin2017, zeh1970} can be shielded, strictly speaking the subsystems of the universe are all open systems.\footnote{Of course, the mere fact that a given system happens to be a subsystem of an entangled system does not imply that its dynamics are non-unitary. Our point here is only that such subsystems are open systems, and that the dynamics of open systems are non-unitary in general.} It is true, of course, that in many cases one can treat a given system of interest as effectively isolated. And even when one cannot do so one can still in many cases model the dynamics of the system, $\mathcal{S}$, by first modeling the dynamics of the larger dynamically coupled closed system, $\mathcal{S+E}$, and then abstracting away from $\mathcal{E}$'s degrees of freedom. But one should not forget that it is the dynamics of $\mathcal{S}$, not the dynamics of $\mathcal{S+E}$, that we should consider ourselves to have successfully described when we do this (as $\mathcal{E}$ will, as we mentioned in Section \ref{s:intro}, typically be highly idealized). The basis of the empirical success of ST, in other words, arguably lies in the way that it effectively describes the dynamics of open systems.\footnote{This argument is our gloss on a point suggested in discussion by Wayne Myrvold.} Since open systems are represented in ST by density operators that evolve non-unitarily in general, there seems to us to be a clear \emph{prima facie} empirical motivation to, without necessarily modifying the quantum-mechanical formalism, formulate a more general theoretical framework in which that is how the fundamental dynamics of systems are most generally described.\footnote{One might object to this that we have not considered whether the simplicity (or other virtues) of the physical laws used in an explanation might play a role in their confirmation. This is an interesting issue, which we discuss in more detail in \citep[sec. 4.3]{cuffaroHartmannOpenSystemsView}, with the upshot being that we can find no way to cash out what we there call the relation of `explanatory fundamentality' in terms of theories that are 'simpler' or 'better' than other theories in a way that gets the right answer to the question, for instance, of whether general relativity or Newtonian gravity is more fundamental. We do not, of course, claim that our arguments are the final word on these rather complicated matters, which is why we qualify the empirical motivation we describe above as only \emph{prima facie}.}

\section{Conclusion}
\label{s:conc}

We have argued that the more general dynamics describable in the theoretical framework of GT can be physically motivated; that there is as much \emph{prima facie} empirical support for GT as there is for ST; and that GT may be as much in the spirit of the Everett interpretation as ST is. There might, in short, be little reason for an Everettian not to embrace GT, and the more general theoretical landscape that it allows one to explore, as the proper conceptual space with which to engage in foundational and philosophical investigations (and speculations) regarding the quantum world.

\vspace{6pt} 



\authorcontributions{Conceptualization, M.E.C. and S.H.; Methodology, M.E.C. and S.H.; Formal analysis, M.E.C. and S.H.; Investigation, M.E.C. and S.H.; Writing---original draft, M.E.C. and S.H.; Writing---review \& editing, M.E.C. and S.H.. All authors have read and agreed to the published version of the manuscript.}

\funding{This research was funded by the German Research Council (DFG), grant number 468374455, and the Alexander von Humboldt Foundation through an Experienced Researcher Grant to M.E.C.
}




\dataavailability{No data has been used to write this paper which we can make available.} 

\acknowledgments{This paper was presented at the conference, ``The Many-Worlds Interpretation of Quantum Mechanics,'' held in Tel Aviv in October 2022. We thank the organizer, Lev Vaidman, and the participants, in particular Charles B\'edard, Eddy Keming Chen, Erik Curiel, G\'abor Hofer-Szab\'o, Alyssa Ney, Simon Saunders, David Wallace, and Cai Waegell for their helpful feedback. Thanks also to Josh Quirke for discussion, to two anonymous referees for their comments on a previous draft, and to the members of our research group on the philosophy of open systems: James Ladyman, S\'ebastien Rivat, David Sloan, and Karim Th\'ebault.
}

\conflictsofinterest{The authors declare that they have no conflicts of interest.} 

\begin{adjustwidth}{-\extralength}{0cm}
\reftitle{References}

\PublishersNote{}
\end{adjustwidth}

\begin{thebibliography}{999}

\bibitem[Myrvold(2022)]{myrvoldInFreireVolume}
Myrvold, W.C.
\newblock Philosophical Issues Raised by Quantum Theory and its
  Interpretations. In {\em The Oxford Handbook of the History of Quantum
  Interpretations}; Freire, O., Ed.; Oxford University Press: Oxford, UK
  , 2022;
  pp. 53--76.

\bibitem[Janas et~al.(2022)Janas, Cuffaro, and Janssen]{3m2020}
Janas, M.; Cuffaro, M.E.; Janssen, M.
\newblock {\em Understanding Quantum Raffles: Quantum Mechanics on an
  Informational Approach: Structure and Interpretation}; Springer: Berlin/Heidelberg, Germany, 
2022.

\bibitem[Brukner(2017)]{brukner2017}
Brukner, {\v{C}}.
\newblock On the Quantum Measurement Problem. In {\em Quantum [Un] Speakables
  {II}}; Springer: Berlin/Heidelberg, Germany, 2017; pp. 95--117.

\bibitem[Bub(2016)]{bub2016}
Bub, J.
\newblock {\em Bananaworld: Quantum Mechanics for Primates}; Oxford University
  Press: Oxford, UK, 2016.


\bibitem[Bub and Pitowsky(2010)]{bub-pitowsky2010}
Bub, J.; Pitowsky, I.
\newblock Two Dogmas About Quantum Mechanics. In {\em Many Worlds? {E}verett,
  Quantum Theory, and Reality}; Saunders, S., Barrett, J., Kent, A., Wallace,
  D., Eds.; Oxford University Press: Oxford,  UK, 2010; pp. 433--459.

\bibitem[Demopoulos(2022)]{demopoulosOnTheories}
Demopoulos, W.
\newblock {\em On Theories}; Harvard University Press: Cambridge, MA, USA,  2022.

\bibitem[Fuchs et~al.(2014)Fuchs, Mermin, and Schack]{fuchsEtAl2014}
Fuchs, C.A.; Mermin, N.D.; Schack, R.
\newblock An introduction to {QBism} with an application to the locality of
  quantum mechanics.
\newblock {\em Am. J. Phys.} {\bf 2014}, {\em 82},~749--754.

\bibitem[Healey(2017)]{healey2017}
Healey, R.
\newblock {\em The Quantum Revolution in Philosophy}; Oxford University Press:
  Oxford, UK, 2017.

\bibitem[Rovelli(2021)]{rovelli2021}
Rovelli, C.
\newblock {\em Helgoland: Making Sense of the Quantum Revolution}; Riverhead
  Books: New York, NY, USA, 2021.

\bibitem[Cuffaro and Hartmann(2021)]{cuffaroHartmannOpenSystemsView}
Cuffaro, M.E.; Hartmann, S.
\newblock The Open Systems View. \emph{arXiv} \textbf{2021},
\newblock {arXiv:2112.11095v1}.

\bibitem[Nielsen and Chuang(2000)]{nielsenChuang2000}
Nielsen, M.A.; Chuang, I.L.
\newblock {\em Quantum Computation and Quantum Information}; {Cambridge
  University Press}: Cambridge, UK, 2000.

\bibitem[Wallace(2019)]{wallace2019}
Wallace, D.
\newblock On the Plurality of Quantum Theories: Quantum Theory as a Framework,
  and its Implications for the Quantum Measurement Problem. In {\em Realism and
  the Quantum}; French, S., Saatsi, J., Eds.; Oxford University Press: Oxford, UK,
  2019; pp. 78--102.

\bibitem[Gleason(1957)]{gleason1957}
Gleason, A.M.
\newblock Measures on the Closed Subspaces of a {H}ilbert Space.
\newblock {\em J. Math. Mech.} {\bf 1957}, {\em
  6},~885--893.

\bibitem[Busch(2003)]{busch2003}
Busch, P.
\newblock Quantum States and Generalized Observables: A Simple Proof of
  {G}leason's Theorem.
\newblock {\em Phys. Rev. Lett.} {\bf 2003}, {\em 91},~120403.

\bibitem[Sudarshan et~al.(1961)Sudarshan, Mathews, and Rau]{sudarshan1961}
Sudarshan, E.C.G.; Mathews, P.M.; Rau, J.
\newblock Stochastic Dynamics of Quantum-Mechanical Systems.
\newblock {\em Phys. Rev.} {\bf 1961}, {\em 121},~920--924.

\bibitem[Jordan and Sudarshan(1961)]{jordan1961}
Jordan, T.F.; Sudarshan, E.C.G.
\newblock Dynamical Mappings of Density Operators in Quantum Mechanics.
\newblock {\em J. Math. Phys.} {\bf 1961}, {\em 2},~772--775.

\bibitem[Jordan et~al.(2004)Jordan, Shaji, and Sudarshan]{jordan2004}
Jordan, T.F.; Shaji, A.; Sudarshan, E.C.G.
\newblock Dynamics of Initially Entangled Open Quantum Systems.
\newblock {\em Phys. Rev. A} {\bf 2004}, {\em 70},~052110.

\bibitem[Shaji and Sudarshan(2005)]{shaji2005}
Shaji, A.; Sudarshan, E.C.G.
\newblock Who's Afraid of Not Completely Positive Maps?
\newblock {\em Phys. Lett. A} {\bf 2005}, {\em 341},~48--54.

\bibitem[Dominy et~al.(2016)Dominy, Shabani, and Lidar]{dominyEtAl2016}
Dominy, J.M.; Shabani, A.; Lidar, D.A.
\newblock A General Framework for Complete Positivity.
\newblock {\em Quantum Inf. Process.} {\bf 2016}, {\em 15},~465--494.

\bibitem[Breuer and Petruccione(2007)]{breuer2007}
Breuer, H.P.; Petruccione, F.
\newblock {\em The Theory of Open Quantum Systems}; Oxford University Press:
  Oxford,  2007.

\bibitem[Pechukas(1994)]{pechukas1994}
Pechukas, P.
\newblock Reduced Dynamics Need Not Be Completely Positive.
\newblock {\em Phys. Rev. Lett.} {\bf 1994}, {\em 73},~1060--1062.

\bibitem[Cuffaro and Myrvold(2013)]{cuffaroMyrvold2013}
Cuffaro, M.E.; Myrvold, W.C.
\newblock On the Debate Concerning the Proper Characterisation of Quantum
  Dynamical Evolution.
\newblock {\em Philos. Sci.} {\bf 2013}, {\em 80},~1125--1136.

\bibitem[Stinespring(1955)]{stinespring1955}
Stinespring, W.F.
\newblock Positive Functions on ${C}^*$-algebras.
\newblock {\em Proc. Am. Math. Soc.} {\bf 1955},
  {\em 6},~211.

\bibitem[Raggio and Primas(1982)]{raggio1982}
Raggio, G.A.; Primas, H.
\newblock Remarks on ``{On} Completely Positive Maps in Generalized Quantum
  Dynamics''.
\newblock {\em Found. Phys.} {\bf 1982}, {\em 12},~433--435.

\bibitem[Saunders(2011)]{saunders2011}
Saunders, S.
\newblock Chance in the Everett Interpretation. In {\em Many Worlds? {E}verett,
  Quantum Theory, and Reality}; Saunders, S.; Barrett, J.; Kent, A.; Wallace,
  D., Eds.; Oxford University Press: Oxford, UK, 2011; pp. 181--205.

\bibitem[Chen(2018)]{chenTimeAsymmetricBJPS}
Chen, E.K.
\newblock Quantum Mechanics in a Time-Asymmetric Universe: On the Nature of the
  Initial Quantum State.
\newblock {\em  Br. J. Philos. Sci.} {\bf 2018},
\newblock \emph{in press}.

\bibitem[Chen(2020{\natexlab{a}})]{chenNatureOfPhysicalLaws}
Chen, E.K.
\newblock The Past Hypothesis and the Nature of Physical Laws. In {\em Time's
  Arrows and the Probability Structure of the World}; Loewer, B., Winsberg, E.,
  Weslake, B., Eds.; Harvard University Press: Cambridge, MA, USA, 2020.

\bibitem[Chen(2020{\natexlab{b}})]{chen2020}
Chen, E.K.
\newblock Time’s Arrow in a Quantum Universe: On the Status of Statistical
  Mechanical Probabilities. In {\em Statistical Mechanics and Scientific
  Explanation: Determinism, Indeterminism and Laws of Nature}; Allori, V., Ed.;
  World Scientific: Jersey City, NJ, USA,  2020; pp. 479--515.

\bibitem[Maroney(2008)]{maroneyGibbsPhysicalBasis}
Maroney, O.J.E.
\newblock The Physical Basis of the {G}ibbs-von {N}eumann Entropy. \emph{arXiv} \textbf{2008},
\newblock {arXiv:quant-ph/0701127v2}.

\bibitem[Robertson(2020)]{robertsonHolyGrail}
Robertson, K.
\newblock In Search of the Holy Grail: How to Reduce the Second Law of
  Thermodynamics.
\newblock {\em The British Journal for the Philosophy of Science} {\bf 2020}.
\newblock In press.

\bibitem[Wallace and Timpson(2010)]{wallaceTimpson2010}
Wallace, D.; Timpson, C.G.
\newblock Quantum Mechanics on Spacetime {I}: Spacetime State Realism.
\newblock {\em  Br. J. Philos. Sci.} {\bf 2010},
  {\em 61},~697--727.

\bibitem[Wallace(2012)]{wallace2012}
Wallace, D.
\newblock {\em The Emergent Multiverse}; Oxford University Press: Oxford, UK, 2012.

\bibitem[Wallace(2020)]{wallace2020a}
Wallace, D.
\newblock Why Black Hole Information Loss Is Paradoxical. In {\em Beyond
  Spacetime: The Foundations of Quantum Gravity}; Huggett, N., Matsubara, K.,
  W\"uthrich, C., Eds.; Cambridge University Press: Cambridge, UK, 2020; pp.
  209--236.

\bibitem[Smeenk and Ellis(2017)]{smeenkEllisCosmologySEP}
Smeenk, C.; Ellis, G.
\newblock Philosophy of Cosmology. In {\em The {S}tanford Encyclopedia of
  Philosophy}, Zalta, E.N., Ed.; Metaphysics Research Lab,
  {S}tanford University: Stanford, CA USA, 2017.

\bibitem[Sloan(2021)]{sloan2021}
Sloan, D.
\newblock New Action for Cosmology.
\newblock {\em Phys. Rev. D} {\bf 2021}, {\em 103},~043524.

\bibitem[Hawking(1976)]{hawking1976b}
Hawking, S.W.
\newblock Breakdown of Predictability in Gravitational Collapse.
\newblock {\em Phys. Rev. D} {\bf 1976}, {\em 14},~2460--2473.

\bibitem[Giddings(2013)]{giddings2013}
Giddings, S.B.
\newblock Black Holes, Quantum Information, and the Foundations of Physics.
\newblock {\em Phys. Today} {\bf 2013}, {\em 66},~30--35.

\bibitem[Page(1983)]{pageUniverseOpenSystem}
Page, D.N.
\newblock Is Our Universe an Open System? In {\em Proceedings of the Third
  Marcel Grossmann Meeting on General Relativity}; Ning, H., Ed.;
  North-Holland: Amsterdam, The Netherlands, 1983; pp. 1153--1155.

\bibitem[Gryb and Sloan(2021)]{grybSloanScaleSurplus}
Gryb, S.; Sloan, D.
\newblock When Scale is Surplus. \emph{arXiv} \textbf{2021},
\newblock {arXiv:2103.07384v2}.

\bibitem[Sloan(2018)]{sloan2018}
Sloan, D.
\newblock Dynamical similarity.
\newblock {\em Phys. Rev. D} {\bf 2018}, {\em 97},~123541.

\bibitem[Castro-Ruiz et~al.(2020)Castro-Ruiz, Giacomini, Belenchia, and
  Brukner]{castroEtAl2020}
Castro-Ruiz, E.; Giacomini, F.; Belenchia, A.; Brukner, {\v{C}}.
\newblock Quantum Clocks and the Temporal Localisability of Events in the
  Presence of Gravitating Quantum Systems.
\newblock {\em Nat. Commun.} {\bf 2020}, {\em 11},~2672.

\bibitem[Giacomini et~al.(2019)Giacomini, Castro-Ruiz, and
  Brukner]{giacominiEtAl2019}
Giacomini, F.; Castro-Ruiz, E.; Brukner, {\v{C}}.
\newblock Quantum Mechanics and the Covariance of Physical Laws in Quantum
  Reference Frames.
\newblock {\em Nat. Commun.} {\bf 2019}, {\em 10},~494.

\bibitem[Oriti(2021)]{oritiTimeEmergence}
Oriti, D.
\newblock The Complex Timeless Emergence of Time in Quantum Gravity. In {\em
  Time and Science}; Harris, P., Lestienne, R., Eds.; World Scientific: Jersey City, NJ, USA, 2021.
  
  
\bibitem[Zeh(2007)]{zeh2007}
Zeh, H.D.
\newblock {\em The Physical Basis of the Direction of Time}, 5th ed.; Springer:
  Berlin/Heidelberg, Germany, 2007.

\bibitem[Herbst et~al.(2015)Herbst, Scheidl, Fink, Handsteiner, Wittmann,
  Ursin, and Zeilinger]{herbstEtAl2015forQuantumReports}
Herbst, T.; Scheidl, T.; Fink, M.; Handsteiner, J.; Wittmann, B.; Ursin, R.;
  Zeilinger, A.
\newblock Teleportation of Entanglement over 143 km.
\newblock {\em Proc. Natl. Acad. Sci. USA} {\bf 2015},
  {\em 112},~14202--5.

\bibitem[Yin et~al.(2017)]{yin2017}
Yin, J.; Cao, Y.; Li, Y.-H.; Liao, S.-K.; Zhang, L.; Ren, J.-G.; Cai, W.-Q.; Liu, W.-Y.; Li, B.; Daiet, H.;
\newblock Satellite-Based Entanglement Distribution over 1200 kilometers.
\newblock {\em Science} {\bf 2017}, {\em 356},~1140--1144.

\bibitem[Zeh(1970)]{zeh1970}
Zeh, H.D.
\newblock On the Interpretation of Measurement in Quantum Theory.
\newblock {\em Found. Phys.} {\bf 1970}, {\em 1},~342--349.

\end{thebibliography}
\end{document}